\documentclass[showpacs,prl,twocolumn,superscriptaddress]{revtex4}
\usepackage{graphicx}
\usepackage{subfigure}

\newcommand{\be}{\begin{equation}}
\newcommand{\ee}{\end{equation}}
\newcommand{\ba}{\begin{eqnarray}}
\newcommand{\ea}{\end{eqnarray}}
\newcommand{\bd}{\begin{description}}
\newcommand{\ed}{\end{description}}

\renewcommand{\iota}{{\bf 1}}
\def\rellow#1#2{Mathrel{Mathop{\kern 0pt #1}\limits_{#2}}}

\begin{document}

\title{Model for Dissipative Highly Nonlinear Waves in Dry Granular Systems}
\author{Lautaro Vergara}
\email{lautaro.vergara@usach.cl}
\affiliation{Departamento de F\'{\i}sica, Universidad de Santiago de Chile, USACH,
Casilla 307, Santiago 2, Chile }
\date{today}

\begin{abstract}
A model is presented for the characterization of dissipative
effects on highly nonlinear waves in one-dimensional dry granular
media. The model includes three terms: Hertzian, viscoelastic and
a term proportional to the square of the relative velocity of
particles. The model outcomes are confronted with different
experiments where the granular system is subject to several
constraints for different materials. Excellent qualitative and
quantitative agreement between theory and experiments is found.
\end{abstract}

\maketitle

%\bigskip

There exist a considerable number of engineering applications
where surfaces are subjected to contact loading, with stress
applied over small areas. Nevertheless, it is a complex matter to
understand the nature of the contact of solid surfaces. A first
step towards this understanding started with the work by Hertz
\cite{hertz}. Using potential theory, he developed a model for the
normal contact on non-conforming bodies of elliptical shapes,
under several assumptions that simplified the problem. One of the
hypotheses on which his work was based is that the contact between
solids is purely elastic.

A place where Hertz theory finds application is in the study of
granular matter. This kind of matter is present everywhere in
Nature and has wide practical importance (see e.g.,
\cite{examples}). Hertz theory has been useful in understanding
many aspects of granular matter \cite{examples}, \cite{granular},
\cite{Fauve}, but its applicability is limited because in many
situations the impact between grains is such that energy
dissipating phenomena become relevant. Such phenomena, involving
for example elasto-plastic and viscoelastic behavior
\cite{Johnson}, are so complex that it is hard to think of a
closed form force law that may describe them all at once.

However, energy loss can be measured experimentally, at the
macroscopic level, by measuring the coefficient of restitution,
which has been observed to decrease with the normal component of
the relative impact velocity \cite{COR}. In this case, Hertz
theory fails to reproduce the behavior of the coefficient of
restitution. In \cite{K-K} and \cite{Brilliantov}, the authors
have developed a quasi-static approximation to calculate the
normal force acting between colliding particles, assuming a
viscoelastic force (see also \cite{Zener} where this force law
appears in a related context, and \cite{Morgado} for a
first-principles derivation of the viscoelastic term). Within this
approach, where the force acting between beads is a combination of
Hertz and viscoelastic terms, theory and experiment agree for the
behavior of the coefficient of restitution with velocity
\cite{Brilliantov2}.

In this Letter, a model that combines Hertz theory, a viscoelastic
force as in \cite{K-K} and \cite{Brilliantov}, and a force
proportional to the square of the relative velocities of beads is
presented. It is shown that this model reproduces in an excellent
way the effect of dissipation on a solitary wave in stainless
steel, brass and polytetraflouroethylene (PTFE) obtained by Daraio
et al. \cite{Daraio}. It also coincides very well with the
behavior of solitary wave trains in a column of stainless steel
beads \cite{DNHJ2} and with the description of incident and
reflected solitary waves in a column of PTFE balls \cite{DNHJ}. To
simulate such systems, I follow as closely as possible the
experimental setup, including the reduction by less than $5.5\%$
in mass of beads with inserted piezosensors.

Let $x_{i}(t)$ represent the displacement of the center of the
$i$-th sphere, of mass $m_i$, from its initial equilibrium
position. The equations of motion that describe the dynamics of
$N$ beads, inclined by an angle $\alpha$, in a gravitational field
are:

\begin{eqnarray}
m_{i}\ddot{x}_{i} &=&K_{i-1,i}\, \delta
_{i-1}^{3/2}-K_{i,i+1}\,\delta _{i}^{3/2}
\nonumber \\
&&+\,\frac{3A}{2}\left\{ \sqrt{\delta _{i-1}}\,\dot{\delta}_{i-1}-\sqrt{%
\delta _{i}}\,\dot{\delta}_{i}\right\}   \nonumber \\
&&+\,B\left\{ \theta_{i-1}\,(sign[\dot{\delta}%
_{i-1}]\,\dot{\delta}_{i-1})^{2}-\theta_{i} \,sign[%
\dot{\delta}_{i}]\,\dot{\delta}_{i}^{2}\right\}   \nonumber \\
&&+\,m_{i}\,g\,\sin [\alpha ],  \label{uno}
\end{eqnarray}
with $i=2,...,N-1$. As known, the equations of motion for the
first and last beads differ from Eq. (\ref{uno}) and are thus not
written down here. The notation is as follows:
$\dot{\delta}_{i}=\dot{x}_{i}-\dot{x}_{i+1}$ is the relative
velocity of beads $i$ and $i+1$. The overlap between adjacent
beads is $\delta _{i}=\max \{\Delta _{i,i+1}-(x_{i+1}-x_{i}),0\}$,
ensuring that the spheres interact only when in contact. For the
same reason a step-function in the third term has been included,
that is, $\theta_{i}=\theta[\Delta _{i,i+1}-(x_{i+1}-x_{i})]$.
$\Delta _{i,i+1}=(g\,\sin [\alpha ]\,i\,m_{i}/K_{i,i+1})^{2/3}$
appears from the pre-compression due to the gravitational
interaction. The expression for the Hertz coupling $K_{i,j}$
between beads $i$ and $j$ is well known and depends on radii,
Young moduli and Poisson ratio of beads \cite{Johnson}. Although
the form of parameter $A$ is known for a binary collision
\cite{Brilliantov}, it is used here as a free parameter, like $B$
itself. The set of Eqs. (\ref{uno}) is solved by using an explicit
Runge-Kutta method of the 5th order with an embedded error
estimator, from {\it{Mathematica}}.

Our explanation for our Ansatz is as follows: after the impact,
the dynamics becomes a multi-impact problem; this produces that
the relative velocity of beads $i$ and $i+1$, $\dot{\delta}_{i}$,
may change from $\dot{\delta}_{i}<0$, related to an expansion phase, to $\dot{%
\delta}_{i}>0$, corresponding to a compressional phase. The same
happens for the relative velocity of beads $i-1$ and $i$,
$\dot{\delta}_{i-1}$. Therefore, by fixing our attention on one
bead in the chain, say $i-th$, one has a particle between two
moving walls (beads $i$ and $i+1$) and then its dynamics depends
on the dynamics of both constraints. It is worthwhile to mention
that a term proportional to the square of the velocity was
introduced by P\"{o}schl \cite{Theodor} as an attempt to extend
the Hertz theory to plastic bodies. Also notice that if beads
$i-1$, $i$ and $i+1$ are all in contact at a given instant, one
can
easily observe that the third term can be generically written as%
\begin{equation}
\dot{x}_{i-1}^{2}+2\,\epsilon\, \dot{x}_{i}^{2}+\kappa\,
\dot{x}_{i+1}^{2}+2\,\eta\, \dot{x}_{i}\left(
\dot{x}_{i-1}-\sigma\, \dot{x}_{i+1}\right), \end{equation} where
the constants take the values $\epsilon =0,1$, $\kappa =\pm 1$,
$\eta =\pm 1$, $\sigma =\pm 1$,
depending on the sign of the relative velocities $\dot{\delta}_{i}$ and $%
\dot{\delta}_{i-1}$. Observe that the force term included in this
Letter is a combination of two terms, only one of them being
dissipative \footnote{In the term to the right, $sign[\dot{\delta}%
_{i-1}]$ multiplies only the first term of $\dot{\delta}_{i-1}$,
i.e. $\dot{x}_{i-1}$}.

In Ref. \cite{Daraio}, the effect of dissipation on the behavior
of solitary waves was clearly shown. Beads made of stainless
steel, brass and PTFE, and a wall of aluminium were used in the
experiment. Their Young modulus and Poisson ratio are: (i)
stainless steel: $E = 193 \times 10^9$ Pa; $\nu = 0.30$; (ii)
brass: $E = 115 \times 10^9$ Pa; $\nu = 0.31$; (iii) PTFE: $E =
1.46 \times 10^9$ Pa; $\nu = 0.46$; and (iv) aluminium: $E = 69
\times 10^9$ Pa; $\nu = 0.33$. Strikers with the same mechanical
properties as beads were used to generate solitary waves; the
force on piezosensors was recorded, and the data presented as
plots of force as a function of time.

In Figure 1, the numerical findings from our model are compared
with those shown in Fig.1(b) of Ref. \cite{Daraio}, that is, for a
chain composed of $N=70$ stainless steel beads, and impactor
velocity $v_1=1.77$ m/s. Sensors are placed in beads 9, 16, 24,
31, 40, 50, 56, and 63. A {\it global} time shift of $\,25 \,\mu
s$ of the extracted data was necessary in order to compare our
findings with the experimental data from Fig.1(b) of Ref.
\cite{Daraio}. This shift possibly can be ascribed to an
experimental time offset. Also, in the data received by the
author, sensors appear in pairs, say sensors 2 and 4 (e.g. beads
16 and 31 for stainless steel). Thus, in order to compare this
experimental data with the simulation, the original experimental
data has been shifted such that the time interval between both
maxima in the data is kept fixed, within an error of $\,0.2 \,\mu
s$. Thus, this shift does not represent a change in the time of
flight, lying within the experimental errors. Figure 1 also shows,
as an inset, the percentage error between experiment and
simulation. One observes that this error is below 20$\%$ for the
maximum amplitude, getting worse when approaching the wings of the
signal, which is not surprising from the experimental point of
view. Figure 2 shows a detail of Figure 1, using original data,
for the force recorded at beads 16 and 56. As one can see, the
agreement between simulations and experiment is quite impressive.
One possible combination of values of the parameters, not
necessarily the optimal one, that gives a quite remarkable
coincidence with experiment is $A=800$ and $B=-1.9$.

%********|*********|*********|*********|*********|*********|*********|****
%********|*********| Fig1:           |*********|*********|****
%********|*********|*********|*********|*********|*********|*********|****

\begin{figure}[t]
\vspace{0.3cm} \centering \includegraphics[width=.43
\textwidth,height=4.5cm]{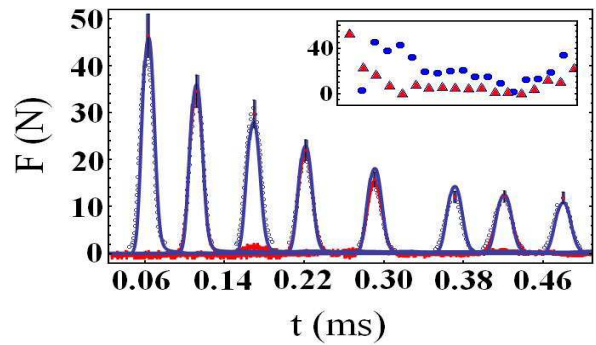} \caption{{The continuous curve
shows our numerical findings. Open circles and error bars are
experimental data extracted from Fig.1(b) of Ref. \cite{Daraio}.
The experimental outcome from sensors 16, 31, 40 and 56 is also
shown. The inset shows the percentage error between experiment and
simulation for beads 16 (triangles) and 40 (circles).}}
\label{panel1}
\end{figure}
%********|*********|*********|*********|*********|*********|*********|****
%********|*********| Fig2:           |*********|*********|****
%********|*********|*********|*********|*********|*********|*********|****
\begin{figure}[t]
\centering \includegraphics[width=.48%
\textwidth,height=2.8cm]{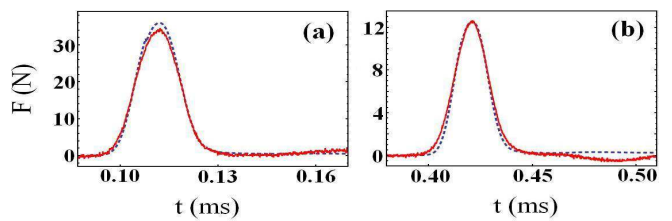} \vspace{-.3cm} \caption{The
figure shows the force as a function of time for beads (a) 16, and
(b) 56, for stainless steel. Dashed lines correspond to the
numerical results.} \label{panel0}
\end{figure}

%********|*********|*********|*********|*********|*********|*********|****
%********|*********| Fig3:           |*********|*********|****
%********|*********|*********|*********|*********|*********|*********|****
\begin{figure}[h]
\vspace{0 cm} \centering \includegraphics[width=.48%
\textwidth,height=2.8cm]{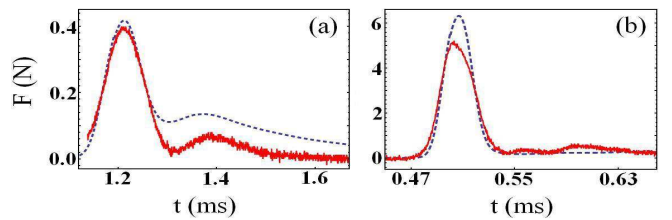} \caption{Plots showing force as
a function of time, for an impact velocity of $v_1=1.55$ m/s, for
(a) PTFE (bead 38), and (b) brass (bead 49).} \label{panel7}
\end{figure}

In Figure 3, a striker with initial velocity $v_1=1.55$ m/s
impacts on a chain of $N=69$ PTFE beads and a chain of $N=61$
brass spheres. Forces are recorded by sensors at beads 38 and 49,
respectively. The numerical values of the parameters that caused
theory and experiment to coincide are $A=100$ and $B=-0.26$, for
PTFE, and $A=770$ and $B=-2.9$, for brass.

In the following, the simulation findings from our model are
compared with those from experiments carried out by Nesterenko et
al. with a column of PTFE beads\cite{DNHJ} lying on a wall made of
brass. There, solitary waves were generated by impacting the
column of 21 beads, each with radius 2.38 mm and mass 0.123 g,
with a PTFE ball with the same characteristics and with a 2.0
$m/s$ impact velocity; the mechanical properties of materials
being the same as above. In Figures 4 (a) and (b), there appear
the incident and reflected perturbations recorded at positions 12
and 16, respectively, while Figure 4 (c) shows the behavior of the
solitary wave at the wall. Because of the lack of detailed
experimental information, global time shifts by 26 $\mu$s and 38
$\mu$s where applied to get coinciding incident perturbations in
figures (a) and (b), respectively, while for figure (c), the shift
made was 31 $\mu$s. Observe that the simulation fits quite well
the experimental data. The same numerical values for the
parameters of the model in case of PTFE are used, i.e. $A=100$ and
$B=-0.26$.
%********|*********|*********|*********|*********|*********|*********|****

%********|*********|*********|*********|*********|*********|*********|****
%********|*********| Fig4:           |*********|*********|****
%********|*********|*********|*********|*********|*********|*********|****
\begin{figure}[t]
\vspace{%
0.0cm}\subfigure{\includegraphics[width=.48\textwidth,height=3.0cm]{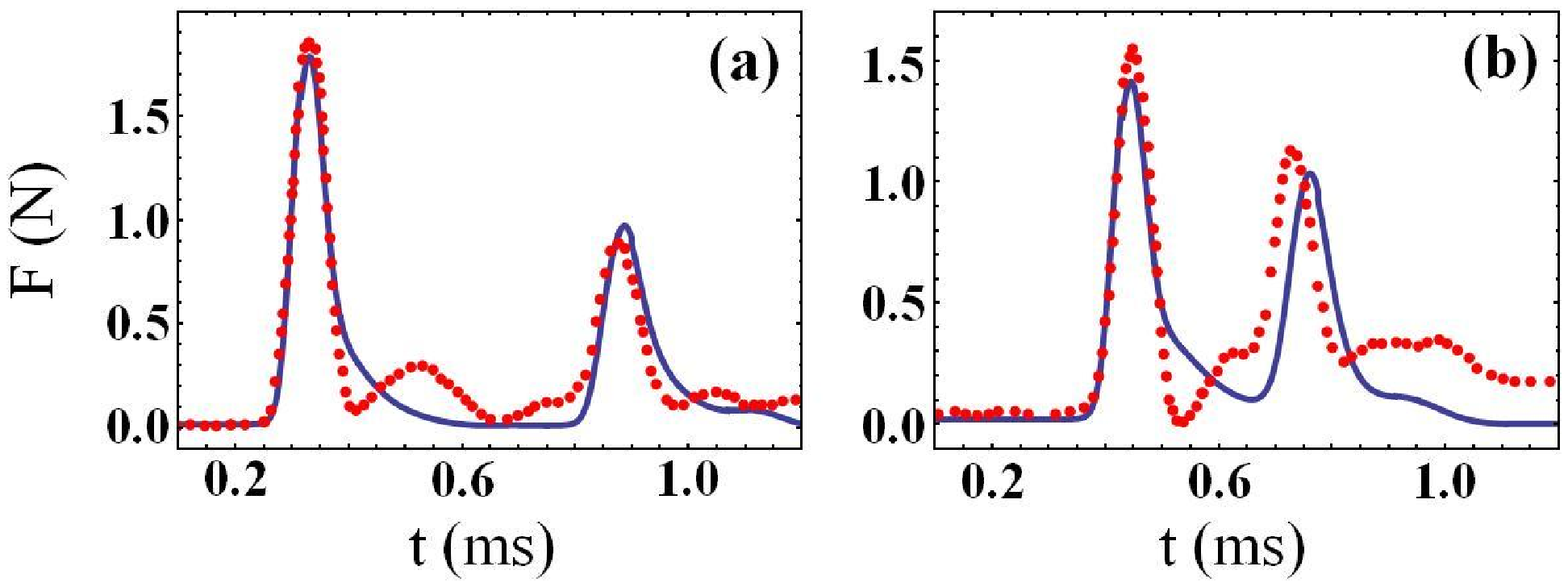}} \vspace{%
0.cm}\subfigure{\includegraphics[width=.476\textwidth,height=2.7cm]{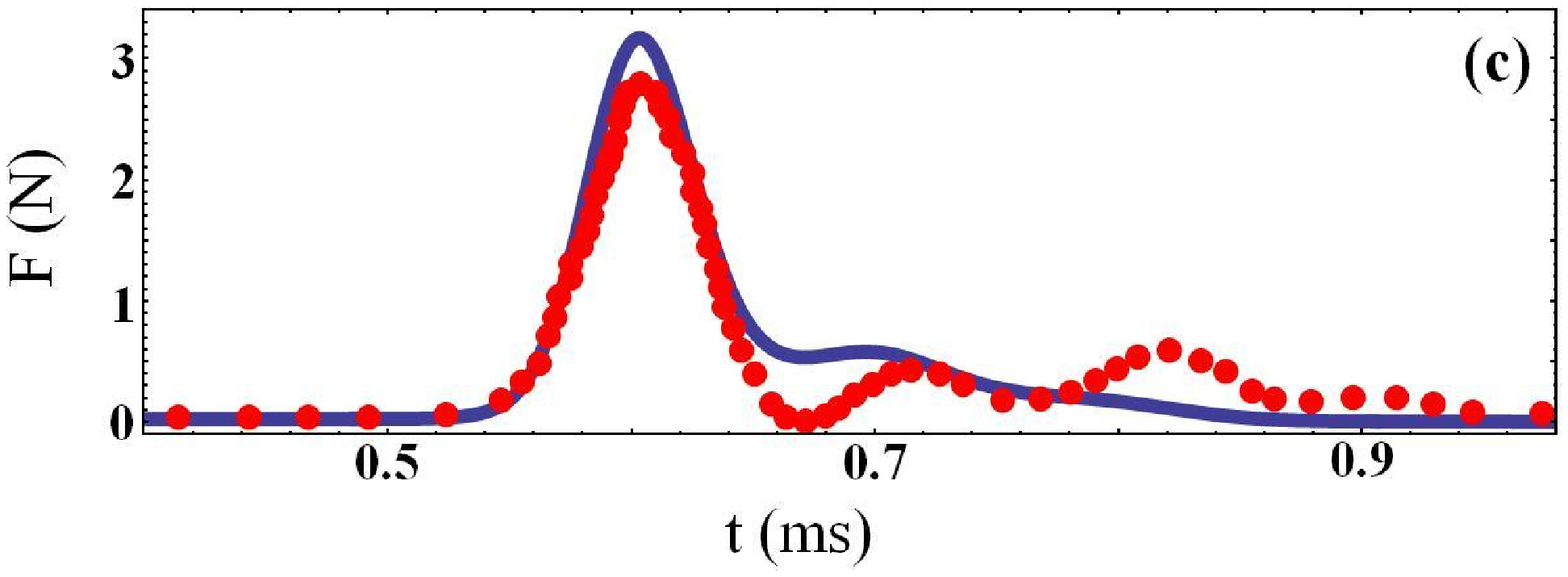}}
\caption{Scattering of highly nonlinear waves off a wall for PTFE
beads at positions: (a) $12$ and (b) $16$. In (c) the sensor is at
the wall. Dots represent experimental data and the solid line the
numerical output.} \label{fig4}
\end{figure}

To end the tests on the validity of the model, the simulation
results are compared with the outcomes from experiments carried
out by Nesterenko et al. with a column of stainless steel beads
\cite{DNHJ2}. The data, extracted from their paper, are used to
show that the model reported here successfully describes the
behavior of highly nonlinear waves on hard walls. In order to
create the nonlinear waves, a column of 21 stainless steel beads,
each with a mass 0.45 g, was struck by a cylindric alumina
impactor, with Young modulus $E = 416 \times 10^9$ Pa and Poisson
ratio $\nu = 0.23$. The cylinder has a mass 1.2 g and a velocity
equal to 0.44 $m/s$. The force is recorded by embedding
piezosensor in beads 12 and 16 (measured from the top of the
column), and at the wall; the piezogauge at the wall was covered
by a brass cover-plate. As expected, because the impactor's mass
is much larger than the mass of beads, trains of highly nonlinear
solitary waves are excited by the impact. Figure 5 shows an
excellent agreement between model and experiment; it must be
stressed that the numerical values for the parameters are the same
as before: $A=800$ and $B=-1.9$. In order to compare experiment
and simulation, the data was shifted 47 $\mu$s for figures (a) and
(b), and 45 $\mu$s for figure (c). The small time difference, of
the order 7 $\mu$s, between the pulse amplitudes from the
simulation and the experiments observed in Figures 4 and 5, (a)
and (b), can be ascribed to the combined properties of both, wall
and piezosensor. If a softer wall is assumed, the time difference
and the amplitude of the reflected wave (and the force at the
wall) can be reduced.

%********|*********|*********|*********|*********|*********|*********|****

%********|*********|*********|*********|*********|*********|*********|****
%********|*********| Fig5:           |*********|*********|****
%********|*********|*********|*********|*********|*********|*********|****
\begin{figure}[t]
\subfigure{\includegraphics[width=.48\textwidth,height=3.0cm]{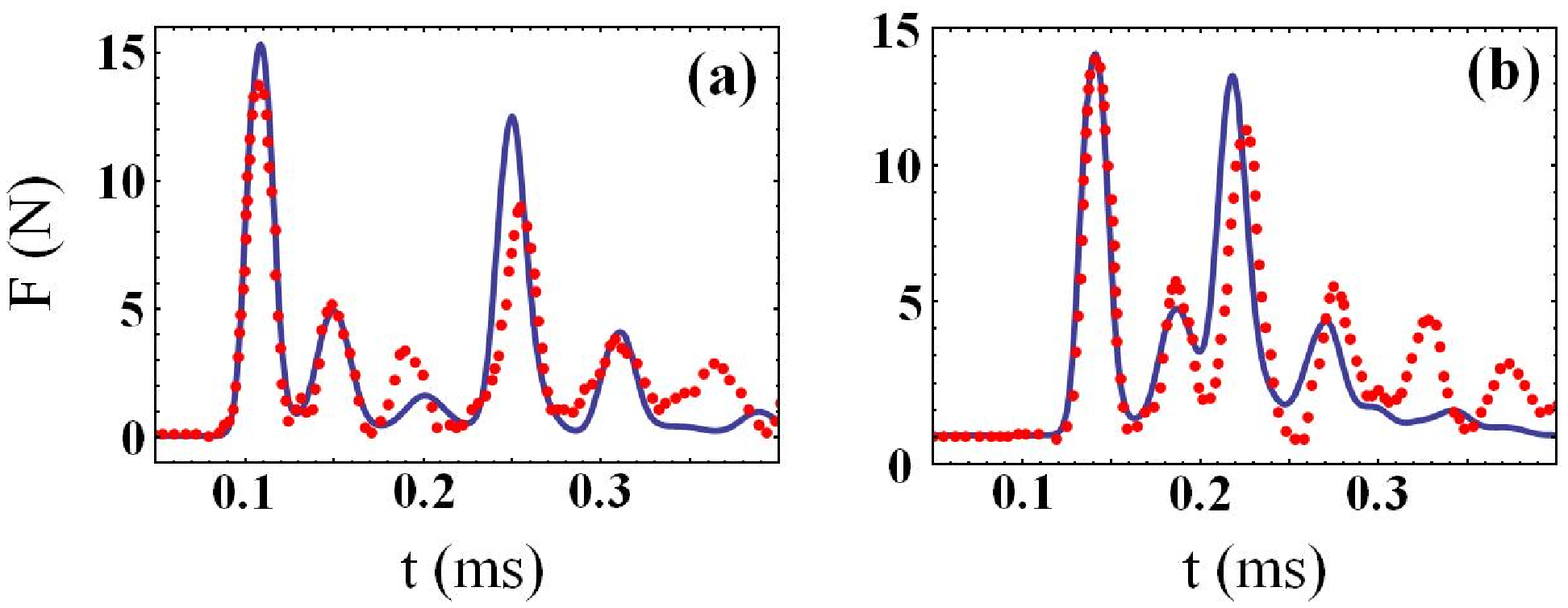}} \vspace{%
0.1cm}\subfigure{\includegraphics[width=.48\textwidth,height=2.8cm]{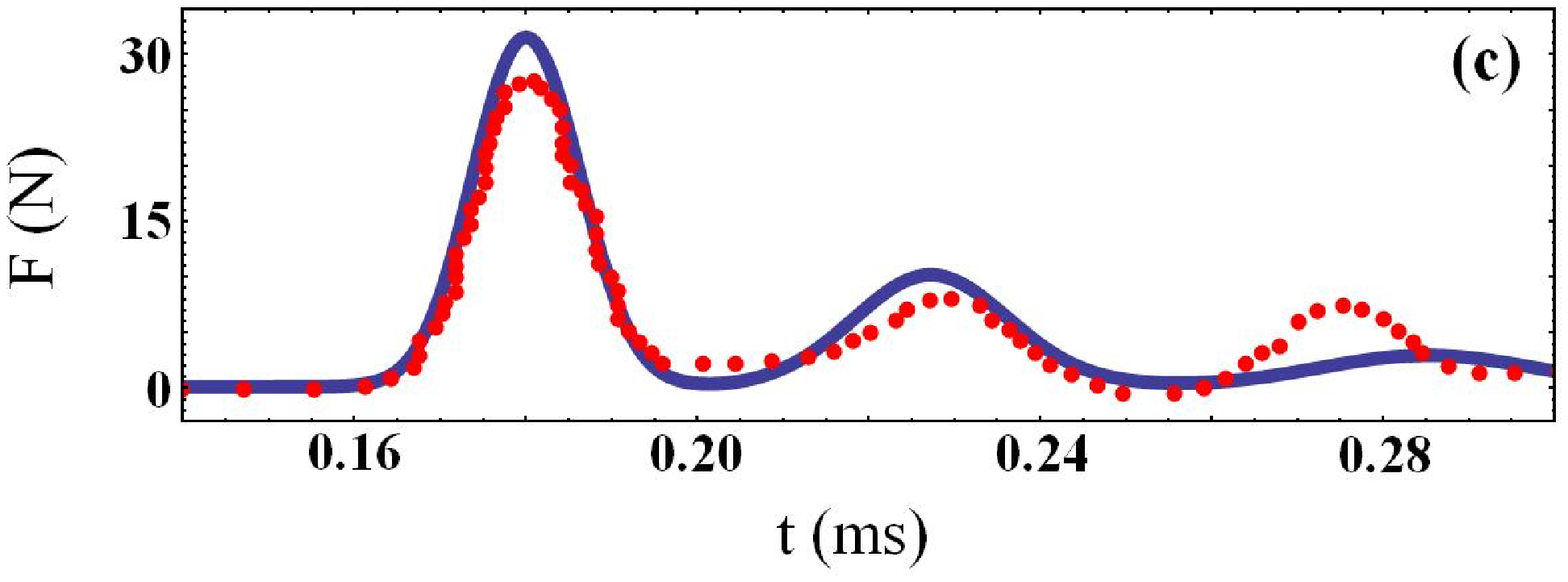}}\vspace{%
-0.3cm} \caption{Scattering of highly nonlinear waves off a wall
for stainless steel beads at positions: (a) $12$, and (b) $16$. In
(c) the sensor is at the wall. Data is shown as dots and numerical
results as a solid line.} \label{fig5}
\end{figure}

Finally, I compare the contribution of the viscoelastic and
velocity-squared terms, using the force on bead 10. The Hertz
interaction is by far the leading term and it is not included
here. In Figure 6, the force on bead 10 is plotted against time;
time units are not made explicit because not all of the plots
correspond to the same time interval. Also, in Figures 6 (a) and 6
(b) force values at the r.h.s. correspond to PTFE. In Figure 6
(a), one observes that both terms give a contribution of the same
order, but in some time intervals each acts in opposition to the
other. A slightly different situation appears in case of the
reflected perturbations, as seen in Figure 6 (b). In this case,
there is a time interval where both force terms reinforce their
contribution. Also observe that the velocity-squared force term is
a continuous, although not smooth, function of time. Nevertheless,
this is not crucial for reproducing the experimental data.

%********|*********|*********|*********|*********|*********|*********|****
%********|*********| Fig6:           |*********|*********|****
%********|*********|*********|*********|*********|*********|*********|****
\begin{figure}[h]
\vspace{0 cm} \centering \includegraphics[width=.48%
\textwidth,height=5.8cm]{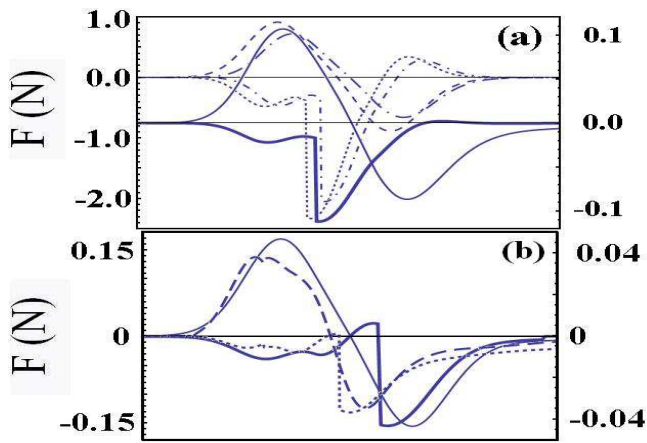}\vspace{%
-0.0cm} \caption{Figure (a) shows viscoelastic (1) and
velocity-squared (2) forces as a function of time on bead 10, for
beads of stainless steel (dashed (1) and dotted (2) lines), brass
(long (1) and short (2) dot-dashed) and PTFE (thin (1) and thick
(2) continuous), with the conditions of Figures 1,2 and 3. Figure
(b) shows the force terms in case of reflected waves, using the
data used in Figures 4 and 5, for stainless steel (thin (1) and
thick (2) continuous) and PTFE (dashed (1) and (2) dotted).}
\label{fig6}
\end{figure}

In conclusion, a model that reproduces experimental outcomes for
the behavior of highly nonlinear dissipative waves, for different
materials and under different conditions, in one-dimensional dry
granular media has been found. Excellent qualitative and
quantitative agreement between theory and experiments is found,
within the experimental errors of measurements of either force
amplitude or time of flight, mostly depending on the Hertz
interaction. The force term added in this Letter, proportional to
the square of the relative velocity of particles, completes a
previous physical model composed of Hertz and viscoelastic
interactions which, without this term, is unable to reproduce
experimental outcomes for the behavior of dissipative highly
nonlinear waves. The new force term originates from multiple
impacts within the system and is composed of two parts, one of
which is dissipative. The model is economical, requiring only two
parameters that depend on the mechanical properties of beads, one
of them being well known and with clear physical origin. In
addition, the parameter values do not need to be changed in order
to fit simulations with experiments carried out under different
conditions, and are found through a single fit procedure. Of
course, if one is looking for the optimal set of parameters, one
should perform a more complete analysis, like the one done in
\cite{Daraio}. The model is, of course, not universally valid and,
for example, is useless for describing the generation of
oscillatory "shock" waves in soft materials as PTFE.

The author is indebted to Prof. C. Daraio for discussions, for
giving details about the experimental setup of Ref. \cite{Daraio}
and for kindly providing most of the data shown in Figs. 1 to 3.
Useful discussions, criticisms and insight about experimental
matters are deeply acknowledged to Prof. R. Labb\'{e}. Useful
suggestions and constructive criticism are acknowledged to Prof.
S. Fauve. This work was partially supported by Fondecyt, Grant No.
1085043.

\end{document}